\documentclass{anstrans}
\title{How accurately we know the standard \CF(sf)~neutron multiplicity?}
\author{R. Capote,$^{*}$ and D. Neudecker$^{\dagger}$}

\institute{
$^{*}$ NAPC--Nuclear Data Section, International Atomic Energy Agency, Vienna, Austria, r.capotenoy@iaea.org
\and
$^{\dagger}$Los Alamos National Laboratory, Los Alamos, NM 87545, USA}

\usepackage{graphicx} 
\usepackage{booktabs} 
\usepackage{microtype} 
\usepackage{esvect}
\usepackage[bookmarksnumbered,colorlinks,bookmarks,breaklinks=true,linkbordercolor={1 1 1},linktocpage,citecolor=blue]{hyperref}

\def\ql{``}

\def\qrs{''}
\def\CF{$^{252}$Cf}

\def\etals{\textit{et al.}}
\def\nub{$\overline{\nu}_{tot}$}

\begin{document}
\section{Introduction}
Statistical evaluation methods of nuclear data have been investigated over several decades~\cite{Smith}, but the dramatic increase in computer power over the last decade has allowed development of powerful stochastic nuclear data evaluation methods~\cite{efnudat2010}. Pioneering works of D.L.~Smith on the Unified Monte Carlo (UMC-G) method \cite{UMCG,UMCG1} have been followed by the development of Backward-Forward Monte Carlo~\cite{BFMC}, UMC-B~\cite{UMCB} and Bayesian Monte Carlo~\cite{BMC}. These evaluation techniques rely on the Bayes theorem (except the BFMC) for combining the experimental data with the physical constraints embedded in nuclear reaction models. Another Bayesian technique---the generalized least-square method (GLSQ)---remains the workhorse of nuclear data evaluation, especially when non-model evaluation is used for cases when the available experimental database is comprehensive. One extremely important case is the evaluation of the Neutron Data Standards~\cite{Carlson:2009,Carlson:2018}.

All evaluation techniques rely 
 on the quality and completeness of the input data used in the selected evaluation method, in particular of the experimental data that drive down our nuclear data uncertainties. This is even more important for non-model evaluations as is the case of Neutron Data Standards, where evaluated uncertainties depend exclusively on the assessed uncertainties of the experimental data used as input to the evaluation.


Small uncertainties obtained for the neutron cross-section standards~\cite{Carlson:2009} have been associated with possible missing correlations in the input data, with an incomplete uncertainty budget of the employed experimental database~\cite{Neudecker:2018} or with unrecognized uncertainty sources common to many measurements. While further detailed studies may improve the first two issues, the issue of potential unrecognized uncertainties and correlations between different experiments has long been neglected.
The goal of this short contribution is to address this gap; the selected test-case study is the evaluation of the total neutron multiplicity (\nub) of the \CF(sf) source, which is included in the evaluation of the Thermal Neutron Constants~\cite{Carlson:2018,Pronyaev:2017}.

\section{Experimental database}

The experimental database used by Axton \cite{axt86} in the evaluation of $\overline{\nu}$$_{tot}$ of $^{252}$Cf(sf) is listed in Table 1. Three different experimental methods have been used in those measurements as detailed by Divadeenam and Stehn \cite{Divadeenam:1984}: scintillator detectors, Mn-bath measurements and Boron pile (B-pile) measurements. Some evidence existed that scintillation measurements produce systematically higher \nub~values as seen in Table 1~\cite{Divadeenam:1984}. However, the calculated \CF(sf) \nub~mean value for scintillation detector measurements is equal to $3.7718\pm 0.0169$, which is higher by just 0.2\% than those corresponding to the Mn-bath detection ($3.7643\pm .0271$). For the two existing Boron-pile measurements the derived mean value is 0.5\% lower than those measured by scintillator detectors, which suggest different systematic uncertainties between different measurement techniques. Note that both Divadeenam and Stehn~\cite{Divadeenam:1984} and also Axton~\cite{axt86} assumed that all \CF(sf) multiplicity measurements are \textit{basically independent measurements}. This is the key assumption that we are questioning in this paper.

\begin{table}[!htb]
    \caption{Fifteen data points used to determine the unrecognized systematic uncertainty for the evaluated $\overline{\nu}$$_{tot}$ of $^{252}$Cf(sf).
    References for these data can be found in the report by Axton~\cite{axt86}.}
    \label{tab:252nu}
    \begin{tabular}{l|llll}
        \hline \hline
       N$^{\circ}$& Author&    Year &    Value (Uncert.) & Method \\
        \hline
        1 & Asplund       &    1963 &    3.7910 (1.066\%) & scintil. \\
        2 & Hopkins       &    1963 &    3.7767 (0.838\%) & scintil. \\
        3 & Boldeman      &    1977 &    3.7549 (0.431\%) & scintil. \\
        4 & Zhang         &    1981 &    3.7534 (0.490\%) & scintil. \\
        5 & Spencer       &    1982 &    3.7831 (0.221\%) & scintil. \\
        \hline
        6 & Colvin/Axton* &    1966 &    3.7299 (0.806\%) & Mn-bath  \\
        7 & DeVolpi       & 1970/72 &    3.7507 (0.463\%) & Mn-bath  \\
        8 & White*        &    1968 &    3.8194 (1.033\%) & Mn-bath  \\
        9 & Axton         &    1985 &    3.7547 (0.300\%) & Mn-bath  \\
        10& Bozorgmanesh  &    1977 &    3.7475 (0.580\%) & Mn-bath  \\
        11& Spiegel       &    1981 &    3.7828 (0.759\%) & Mn-bath  \\
        12& Aleksandrov   &    1981 &    3.7618 (0.483\%) & Mn-bath  \\
        13& Smith         &    1984 &    3.7678 (0.303\%) & Mn-bath  \\
        \hline
        14& Colvin/Ullo   &    1965 &    3.7405 (0.438\%) & B-pile  \\
        15& Edwards       &    1982 &    3.7641 (0.711\%) & B-pile  \\
        \hline  \hline
    \end{tabular}
\vspace{-1mm}
\end{table}
The standard value of \nub~for \CF(sf) was derived using the whole database for all fissile nuclei and is equal to $3.764\pm 0.005$~($0.13$\%)~\cite{Pronyaev:2017}. This value is identical to the one derived from the GLSQ fit of the fifteen experiments listed in Table~1, if we assume that all experiments are independent. This is not surprising as measurements of neutron multiplicity of \CF(sf) are among the most accurate neutron measurements featuring the lowest achievable uncertainty. Evaluated mean value and the corresponding standard deviation are represented in Fig.~1 by dashed red lines and compared to all measurements and corresponding uncertainties listed in Table~1. It is observed that the spread of measured mean values is much larger than the standard deviation of 0.13\% derived from a GLSQ fit (red dashed lines). The derived uncertainty band does not overlap with Spencer 1982 measurement (set 5 in Fig.~1) that features the lowest uncertainty.

\begin{figure}[ht] 
  \centering
  \includegraphics[width=\columnwidth]{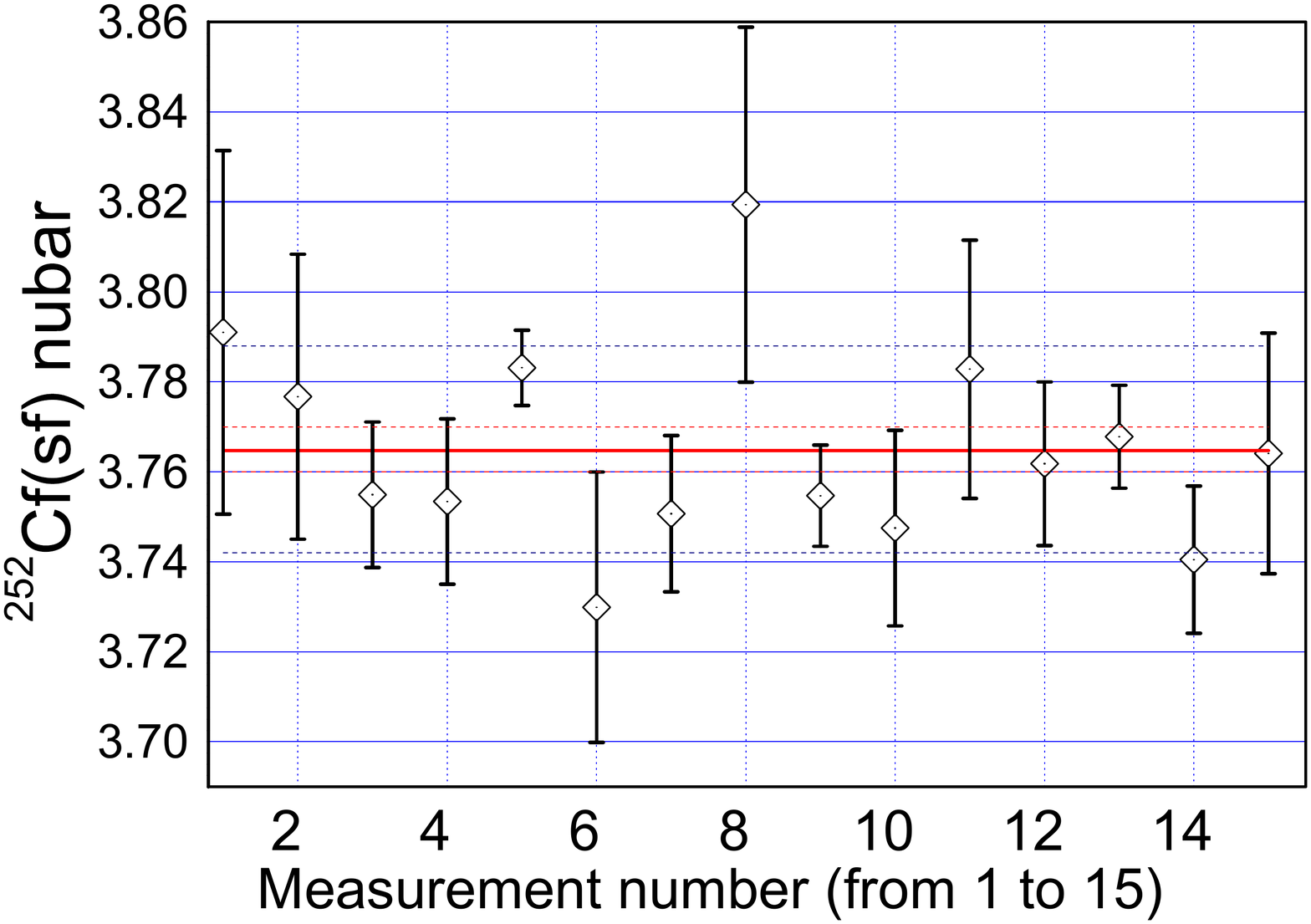}
  \caption{Plot of the fifteen measured points of the $\overline{\nu}$$_{tot}$ of $^{252}$Cf(sf) with uncertainties. Red dashed line denoted the mean value and 1-sigma uncertainty band from the GLSQ fit. Blue dashed lines show the uncertainty bands if USU is added.}
  \label{fig:voltage}
\end{figure}

On the other side, if we calculate the \textit{un-weighted} mean value and the standard deviation of the fifteen \textit{un-weighted} measured data listed in Table~1 we obtain $3.765 \pm 0.023$ (0.6\%). The un-weighted fit produces a larger standard deviation of 0.6\% (blue dashed lines in Fig.~1) which better reflects the observed spread in measured values. This larger uncertainty is strongly reduced by a factor of $1/\sqrt{15}$ in the \textit{weighted} least-square fit due to the assumption that all $N=15$ measurements are practically independent. Indeed, the derived GLSQ uncertainty of $0.005\approx 0.023/\sqrt{15}$. No matter how detailed and complete the uncertainty budget of individual measurements listed in Table~1 be, as long as those measurements are assumed to be independent, the resulting evaluated uncertainty will be too small due to sheer number of input data sets (N=15)~\cite{Neudecker:2013}.

What could be the origin of the correlations that may keep the evaluated uncertainty above some minimum threshold, independent of the number of measurements?
Obviously, an unknown systematic uncertainty, or as D.~Rumsfeld defined it, an \textit{unknown unknowns}\footnote{D. Rumsfeld stated: ``Reports that say that something hasn't
happened are always interesting to me, because as we know, there are known knowns; there are things we know we know. We also
know there are known unknowns; that is to say we know there are some things we do not know. But there are also unknown
unknowns the ones we don't know we don't know. . . , it is the latter category that tend to be the difficult ones.''}.

The unknown (or unrecognized~\cite{Badikov:1992,Badikov:2003,Gai:2007a}) systematic uncertainty (USU) can de defined as a minimum uncertainty that can be achieved using a given experimental method (or measuring tool). It does not matter how many times the measurements are repeated, it does not matter who is doing the measurement, or where those measurements are carried out. As long as we use the same experimental technique, then we can not get a result with lower uncertainty than USU. An interesting consequence is that we need measurements using new experimental methods as the way of estimating these unknown systematic uncertainties.

Authors of Neutron Standard evaluations \cite{Carlson:2009,Carlson:2018,Badikov:2003} have investigated the unknown systematic uncertainties (USU) based on the unrecognized uncertainty-estimation method~\cite{Badikov:1992,Badikov:2003,Gai:2007a}. Each of the cross sections evaluated (and also the $^{252}$Cf(sf) neutron multiplicity) had the normalization quantities for absolute measurements (or the measurements itself) statistically analyzed to obtain the standard deviation of that distribution. The value of the derived standard deviation was assumed as the normally-distributed type B uncertainty~\cite{JCGM100} that is equal to the USU. An additional assumption was made in the Neutron Standard analysis that the USU is not energy dependent~\cite{Carlson:2018}.

We have demonstrated in the Appendix~\ref{appendix} that this assumption is only valid for a one dimensional case. In other words, once the USU is estimated a new least-square fit is needed, except for the case when only a one-dimensional quantity is evaluated. Therefore, an updated Neutron Standard evaluation may be required. Further work on this particular issue is warranted, but the neutron multiplicity of \CF(sf) is a one dimensional quantity. The USU of the \CF(sf) \nub~estimated in the Neutron Standards \cite{Carlson:2018} was equal to 0.4\% as two outlier values (marked by * in Table 1) were discarded. If we consider all experimental values a larger USU of 0.6~\% is obtained.

The most important consequence of the introduced USU for the \CF(sf)~\nub~evaluation is the fact that the USU also determines full cross-correlations between unrecognized uncertainties of different experiments. This estimate represents a conservative uncertainty assessment.

\section{Conclusions}
The unknown systematic uncertainty (USU) for neutron multiplicity of the \CF(sf) is estimated from the measured values to be 0.6\% if the whole experimental database is used. This value is a conservative estimate of the achievable uncertainty. The estimated USU can be interpreted as a common (unknown) systematic uncertainty for all experiments leading to inter-experiments correlations.

\section{Acknowledgments}
Extensive discussions with V.G. Pronyaev and A. D. Carlson provided significant insight on the use of the unrecognized uncertainty method in the evaluation of Neutron Standards.

\bibliographystyle{ans}

\begin{thebibliography}{9}

\bibitem{Smith} D.L. Smith, ``Probability, Statistics, and Data Uncertainties in Nuclear Science and Technology,'' American Nuclear Society, LaGrange Park, IL, USA (1991).

\bibitem{efnudat2010} R. Capote, D. L. Smith, A. Trkov, \ql Nuclear Data Evaluation Methodology Including Estimates of Covariances,\qrs  \textsc{EPJ Web of Conf. }~\textbf{8}, 04001 (2010).

\bibitem{UMCG} D.L. Smith, \ql A Unified Monte Carlo Approach to Fast Neutron Cross Section Data Evaluation,\qrs~in \textsc{Proceedings of the
8th International Topical Meeting on Nuclear Applications and Utilization of Accelerators}, Pocatello, ID, July 29-Aug 2, 2007, p. 736

\bibitem{UMCG1} R. Capote and D.L. Smith, \ql An Investigation of the Performance of the Unified Monte Carlo Method of Neutron Cross Section Data Evaluation,\qrs
\textsc{Nucl. Data Sheets }~\textbf{109}, 2768 (2008).

\bibitem{BFMC} E. Bauge, P. Dossantos-Uzarralde, \ql Evaluation of the covariance matrix of Pu-239 neutronic cross sections in the continuum using a Backward-Forward Monte Carlo method,\qrs~\textsc{J. Kor. Phys. Soc. }\textbf{59}, 1218 (2011).

\bibitem{UMCB} R. Capote, D.L. Smith, A. Trkov, M. Meghzifene, \textsc{J. ASTM Int. }\textbf{9}, 104115 (2012).

\bibitem{BMC} A.J. Koning, \ql Bayesian Monte Carlo method for nuclear data evaluation,\qrs~  \textsc{EPJ }\textbf{A51}, 184 (2015).


\bibitem{Carlson:2009} A. D. Carlson, V. G. Pronyaev, D. L. Smith \etals, \ql International Evaluation of Neutron Cross Section Standards,\qrs  \textsc{ Nucl. Data Sheets }\textbf{110}, 3215--3324~(2009).

\bibitem{Carlson:2018} A. D. Carlson, V. G. Pronyaev, R. Capote \etals, \ql Evaluation of Neutron Data Standards,\qrs  \textsc{ Nucl. Data Sheets} {\bf 148}, 142--187 (2018).

\bibitem{Neudecker:2018} D. Neudecker, B. Hejnal, F. Tovesson, M. C. White, D. L. Smith, D. Vaughan, and R. Capote, \ql Template for estimating uncertainties of measured neutron-induced fission cross-sections,\qrs presented at Covariance workshop in Aix-en-Provence, October 2017; \textsc{EPJ Web of Conf. }, accepted for publication in 2018.

\bibitem{Pronyaev:2017} V. G. Pronyaev, R. Capote, A. Trkov, G. Noguere, and A. Wallner, \ql New fit of thermal neutron constants (TNC) for $^{233,235}$U, $^{239,241}$Pu and \CF(sf): Microscopic vs. Maxwellian data,\qrs~\textsc{EPJ Web of Conf. }\textbf{146}, 02045 (2017).

\bibitem{axt86} E.J. Axton, ``Evaluation of the thermal constants of $^{233}$U, $^{235}$U, $^{239}$Pu and $^{241}$Pu, and the fission neutron yield of $^{252}$Cf,'' Report \textbf{GE/PH/01/86} (Central Bureau for Nuclear Measurements, Geel 1986).

\bibitem{Divadeenam:1984} M. Divadeenam and J.R. Stehn, \ql A least-squares evaluation of thermal data for fissile nuclei,\qrs  \textsc{ Ann. Nucl. Energy} {\bf 11}, 375 (1984).

\bibitem{Neudecker:2013} D.~Neudecker, R.~Capote and H.~Leeb \ql Impact of model defect and experimental uncertainties on evaluated output,\qrs  \textsc{ Nucl. Instr. Methods A} {\bf 723}, 163--172 (2013).

\bibitem{Badikov:1992}  S.A. Badikov, E.V. Gai, M.A. Guseynov and N.S. Rabotnov, \ql Nuclear data processing, evaluation, transformation and storage with Pad\'e-approximants,\qrs Proc. Int. Conf. Nuclear Data for Science and Technology, J\"ulich, Germany, 1991, ed: S.M. Qaim, 182--187 (1992) Springer-Verlag, Berlin.

\bibitem{Badikov:2003} S.A. Badikov, E.V. Gai, ``Some Sources of the Underestimation of Evaluated Cross-section Uncertainties,'' \textbf{INDC(NDS)-438} (IAEA, Vienna 2003), pp. 117--129. Available online at \url{https://www-nds.iaea.org/publications/indc/indc-nds-0438.pdf}.

\bibitem{Gai:2007a} E.V. Gai, ``On the Problem of Ambiguity of the Evaluated Nuclear Data Uncertainties,''
\textsc{ Vopr. Atom. Nauki i Tech., Ser. Nucl. Constants}, issue 1-2, pp. 45--55 (2007); in Russian.

\bibitem{JCGM100} JCGM 100, \ql Evaluation of measurement data: Guide to the expression of uncertainty in measurement,\qrs~BIPM
publication available at \url{https://www.bipm.org/utils/common/documents/jcgm/JCGM_100_2008_E.pdf} (2008).

\bibitem{ShermanMorrison}
J.~Sherman and W.J.~Morrison, \ql Adjustment of an Inverse Matrix Corresponding to a Change in One Element of a Given Matrix,\qrs {\sc Ann. of Math. Statist.}, {21}, 124--127 (1950).

\end{thebibliography}

\section{Appendix}
\label{appendix}
In the evaluation of Neutron Data Standards~\cite{Carlson:2018}, the USU uncertainty was added a-posteriori to the evaluated uncertainties obtained by the GLSQ fit.
Here, we show that adding the USU a-posteriori to the evaluated mean values rather than a-priori does not change the evaluated mean value and variance if one evaluates one data point only while this does not hold true for evaluating multiple data points.

If only one observable (e.g., the $^{252}$Cf \nub)  is evaluated using a vector of experimental data $r_j$ and an associated covariance matrix with elements $C_{ij}$, GLSQ gives an evaluated mean value $\rho$ and a variance $\mathrm{var}(\rho)$ by:
\begin{align}
\label{eq:GLSQ-1dim}
\rho & = \sum_{ij}(C^{-1})_{ij}r_j/\sum_{ij}(C^{-1})_{ij}, \\
\nonumber
\mathrm{var(\rho)} & = 1.0/\sum_{ij}(C^{-1})_{ij}.
\end{align}
For simplicity, we assume that the original covariance matrix (i.e., without the USU) is diagonal such that $C_{ij}^o=\delta_{ij}\sigma_i^2$ with $\delta_{ii}=1$ and 0 if $i\neq j$.
This assumption was also made for the evaluation of Axton.
If the USU variance $\sigma^2_{USU}$ is added a-posteriori in quadrature to the variance evaluated with $C_{ij}^o$ in Eq.~\eqref{eq:GLSQ-1dim}, the a-posteriori variance and mean value read:
\begin{align}
\label{eq:aposteriori-1dim}
\mathrm{var(\rho)}^{post} & = 1.0/\sum_{i}(\sigma_i^{-2})+\sigma^2_{USU} =\sigma_o^{2}+\sigma^2_{USU} , \\
\nonumber
\rho^{post} & = \sigma_o^{2}\sum_{i}r_i\sigma_i^{-2}.
\end{align}
If $\sigma^2_{USU}$ is applied a-priori to the experimental covariance matrix for Eq.~\eqref{eq:GLSQ-1dim} turns into  $C_{ij}^p=\delta_{ij}\sigma_i^2+\sigma^2_{USU}=C_{ij}^o+\sigma^2_{USU}$.
If one uses the Sherman-Morrison formula~\cite{ShermanMorrison} to invert $C_{ij}^p$ and uses Eq.~\eqref{eq:GLSQ-1dim} to obtain an evaluated mean value and variance, one obtains the same equations as in Eq.~\eqref{eq:aposteriori-1dim}.

If one evaluates multiple observables (i.e., a vector $\vv{\rho}$) based on multiple experiments for each observable stored in a vector $\vv{r}$ and an associated covariance matrix $\mathbf{C}$, Eq.~\eqref{eq:GLSQ-1dim} turns into:
\begin{align}
\label{eq:GLSQ-multidim}
\vv\rho & = \mathrm{\mathbf{Cov}}(\vv\rho)\mathbf S^T\mathbf{C}^{-1}\vv r, \\
\nonumber
\mathrm{\mathbf{Cov}}(\vv\rho) & = \left(\mathbf S^T\mathbf{C}^{-1}\mathbf S\right)^{-1}.
\end{align}
For simplicity, we assume to evaluate two data points $\rho_a$ and $\rho_b$ based on a transposed vector $\vv{r}^T = (r_{a,1},\cdots, r_{a,N},r_{b,1},\cdots, r_{b,N})$ and a  transposed design matrix
\begin{equation}
\nonumber
\mathbf S^T =
\begin{bmatrix}
1 & \cdots 1 & 0 & \cdots & 0\\
0 & \cdots 0 & 1 & \cdots & 1\\
 \end{bmatrix}.
\end{equation}
Again the original covariance matrix $\mathbf C^o$ (i.e., without the USU) is diagonal with variances $\sigma_{a,1}^2,\cdots,\sigma_{a,N}^2,\sigma_{b,1}^2,\cdots,\sigma_{b,N}^2$.
If the USU variance $\sigma^2_{USU}$ is added a-posteriori in quadrature to the covariance matrix evaluated with $C_{ij}^o$ in Eq.~\eqref{eq:GLSQ-multidim}, the a-posteriori covariance and mean values read:

\begin{align}
\label{eq:aposteriori-multidim}
 \vv \rho^{post} =  \begin{bmatrix}
 \sigma_{a,o}^{2} & 0\\
 0 & \sigma_{b,o}^{2}\\
  \end{bmatrix}\begin{bmatrix}
 \sum_i r_{a,i} \sigma_{a,i}^{-2}\\
\sum_i r_{b,i}\sigma_{b,i}^{-2}\\
   \end{bmatrix}, \\
   \nonumber
\mathrm{\mathbf{Cov} }(\vv \rho^{post}) =
\begin{bmatrix}
\sigma_{a,o}^{2}+\sigma^2_{USU}  & \sigma^2_{USU}\\
\sigma^2_{USU} & \sigma_{b,o}^{2}+\sigma^2_{USU}\\
 \end{bmatrix}.
\end{align}

If $\sigma^2_{USU}$ is applied a-priori to the experimental covariance matrix for Eq.~\eqref{eq:GLSQ-multidim}, one adds to all elements  of $\mathbf C^o$ the term $\sigma^2_{USU}$ and obtains $\mathbf{C}^p$.
If one uses the Sherman-Morrison formula~\cite{ShermanMorrison} to invert $\mathbf C^p$ and uses Eq.~\eqref{eq:GLSQ-multidim} to obtain an evaluated mean value and variance, one obtains different mean values and covariances compared to Eq.~\eqref{eq:aposteriori-multidim}, namely:
\begin{align}
\label{eq:apriori-multidim}
 \vv \rho^p & =  \vv \rho^{post}+\begin{bmatrix}
 \sigma^2_{USU}\left(s_a+s_b)\sigma_{a,o}^{2}\sum_i r_{a,i} \sigma_{a,i}^{-2}-r_0\right)\\
\sigma^2_{USU}\left((s_a+s_b)\sigma_{b,o}^{2}\sum_i r_{b,i} \sigma_{b,i}^{-2}-r_0\right)\\
   \end{bmatrix}, \\
   \nonumber
\mathrm{\mathbf{Cov} }(\vv \rho^{p}) & = [1+\sigma^2_{USU}(s_a+s_b)]
\begin{bmatrix}
\sigma_{a,o}^{2}+\sigma^2_{USU}  & \sigma^2_{USU}\\
\sigma^2_{USU} & \sigma_{b,o}^{2}+\sigma^2_{USU}\\
 \end{bmatrix},
\end{align}
with $r_0=\sum_i r_{a,i} \sigma_{a,i}^{-2}+\sum_i r_{b,i} \sigma_{b,i}^{-2}$, $s_a=\sum_i \sigma_{a,i}^{-2}$ and $s_b=\sum_i \sigma_{b,i}^{-2}$.
Hence, the evaluated mean values and covariances are expected to change if $\sigma^2_{USU}$ is applied a-priori to a Neutron Data Standard evaluation of multiple observables (e.g., a cross-section given for multiple incident energies.)


\end{document}